\documentclass[12pt]{article}
\usepackage{graphicx}
\usepackage[square,sort,comma,numbers]{natbib}
\usepackage{mystyle}
\graphicspath{{figs/}}
\newcommand{\EQ}[1]{Eq.~\eqref{#1}}
\newcommand{\FIG}[1]{Fig.~\ref{fig:#1}}
\newcommand{\SEC}[1]{Sec.~\ref{sec:#1}}

\newcommand\pubnumber{} 
\newcommand\pubdate{\today}

\def\glasgow{SUPA, School of Physics and Astronomy\\
University of Glasgow, Glasgow, G12 8QQ, UK}

\def\Title#1{\begin{center} {\Large #1 } \end{center}}
\def\Author#1{\begin{center}{ \sc #1} \end{center}}
\def\Address#1{\begin{center}{ \it #1} \end{center}}

\newcommand\pubblock{\rightline{\begin{tabular}{l} \pubnumber\\
         \pubdate  \end{tabular}}}
\newenvironment{Abstract}{\begin{quotation}  }{\end{quotation}}
\newenvironment{Presented}{\begin{quotation} \begin{center}
             PRESENTED AT\end{center}\bigskip
      \begin{center}\begin{large}}{\end{large}\end{center} \end{quotation}}
\def\Acknowledgements{\bigskip  \bigskip \begin{center} \begin{large}
             \bf ACKNOWLEDGEMENTS \end{large}\end{center}}

\def\beq{\begin{equation}}
\def\eeq#1{\label{#1}\end{equation}}
\def\eeqn{\end{equation}}

\def\beqa{\begin{eqnarray}}
\def\eeqa#1{\label{#1}\end{eqnarray}}
\def\eeqan{\end{eqnarray}}

\let\bar=\overbar

\def\O{{\cal O}}

\def\Dslash{\not{\hbox{\kern-4pt $D$}}}
\def\dslash{\not{\hbox{\kern-2pt $\del$}}}

\def\msb{{\bar{\ssstyle M \kern -1pt S}}}

\begin{document}
\begin{titlepage}
\pubblock

\vfill
\Title{Charm and bottom quark masses on the lattice}
\vfill
\Author{Andrew T.\ Lytle}
\Address{\glasgow}
\vfill
\begin{Abstract}
Lattice determinations of quark mass have made significant
progress in the last few years.
I will review recent advances in calculations of 
charm and bottom mass, which are near to achieving
percent-level precision and with fully controlled systematics.
Precise knowledge of these parameters is
of particular interest for precision Higgs studies at future accelerators.

\end{Abstract}
\vfill
\begin{Presented}
The 7th International Workshop on Charm Physics (CHARM 2015)\\
Detroit, MI, 18-22 May, 2015
\end{Presented}
\vfill
\end{titlepage}
\def\thefootnote{\fnsymbol{footnote}}
\setcounter{footnote}{0}
%

\section{Introduction}


Quark masses are fundamental parameters entering into the definition
of the Standard Model.
Within the Standard Model picture, 
quark masses arise from Yukawa interactions with the Higgs field, and
direct measurements of the Higgs couplings at the LHC are consistent
with Standard Model predictions.
High-precision studies at future accelerators such as the ILC will
measure couplings at the per mil level~\cite{Lepage:2014fla}.  
In order to test the SM at this level,
and to constrain and potentially discriminate between models of new physics
detectable at this level,
it is imperative to determine the quark masses to a corresponding
level of precision.

In recent years, considerable progress has been made in lattice
calculations of quark masses, with groups now quoting charm and
bottom mass values at around the percent or few-percent level.
This is due to increasingly realistic simulations, and new techniques.
State-of-the-art simulations include dynamical $u,d,s$, 
and frequently $c$ quarks, with pion masses reaching their
physical values, and typically at several lattice spacings.
This increased realism translates into increasingly accurate results,
and with fewer systematic errors.
In order to reliably determine quark masses at the sub-percent
level, it is important to have a variety of calculational 
techniques/strategies available, along with
independent determinations from different groups.

The outline of the rest of this article is as follows:
\SEC{2} briefly discusses quark mass parameters in a general context,
and how they are determined in lattice QCD simulations.
In~\SEC{charm} I will discuss recent progress in the charm mass determinations, 
focusing on a promising method using current-current correlators.
\SEC{bottom} will look at strategies and results for
 bottom mass determinations,
and~\SEC{ratios} discusses the important roled played by mass ratios.
\SEC{conclusion} presents a summary and discusses future prospects for these
calculations.

\section{Quark mass and LQCD} \label{sec:2}

Quark masses are scheme and scale dependent quantities and can
be viewed as input parameters that, along with $\alpha_s$,
specify QCD at the Lagrangian level.
These parameters must ultimately be determined from experiment,
but because quarks are confined into hadrons the connection is
necessarily indirect.
In the absence of lattice simulations, one must focus on
experimentally measureable observables which are 
1) sensitive to quark masses and 
2) can be reliably computed in perturbation theory.
One set of observables satisfying these criteria are 
derived from the the R-ratio.
Much effort has gone into calculation of the relevant perturbation series,
which are now known to 
$\text{N}^3$LO~\cite{Chetyrkin:2006xg,Boughezal:2006px,Maier:2009fz}.
As will be discussed in~\SEC{jj-charm}, one promising way to calculate
$m_q$ for heavy quarks via lattice simulations uses the same
perturbative calculations,
but substitutes experimental data with data from LQCD simulations.

Lattice QCD simulations are well suited for mass determinations,
since the mass parameters are simulation inputs controlled by 
the ``experimenter''.  
By changing the input masses, one can directly measure the resultant
change in physical observables.  
In a standard LQCD simulation, one tunes the input
masses in order to reproduce the masses of some low-lying hadrons --
one for each quark in the theory.
In this way one obtains (typically very precise) bare quark masses,
but in the particular lattice regularization one happens to be using.
In order to make contact with a continuum-regularized determination
such as the $\MSbar$ scheme, one needs an additional calculation of
the lattice to $\MSbar$ matching factor.  This can be found using 
lattice perturbation theory or via non-perturbative
renormalization (NPR) techniques.
The ratios of bare quark masses in a given regularization
are however immediately useful, as they are equal
to renormalized mass ratios (up to lattice artifacts).

\section{Charm quark mass} \label{sec:charm}

\subsection{Current-current correlator method.} \label{sec:jj-charm}
The current-current correlator method uses moments of 
Euclidean-time twopoint functions,
\begin{equation}
G(t) = a^6 \sum_{\mathbf{x}} (a m_{0h})^2 
\langle J_5(t, \mathbf{x}) J_5(0, 0) \rangle \, .
\end{equation}
Here $J_5 \equiv \bar{\psi}_h \g_5 \psi_h$ and $a m_{0h}$ is the
bare quark mass parameter in lattice units.
In formalisms with sufficient chiral symmetry, 
the current $J_5$ is absolutely normalized.
The correlator $G(t)$ is UV finite, so that
\begin{equation}
G(t)_{\text{cont}} = G(t)_{\text{latt}} + \O(a^2) \qquad (t \neq 0) \, .
\end{equation}
The correlators $G(t)_{\text{latt}}$ are the same ones used to
compute pseudoscalar masses and decay constants, in which case it is the
large-$t$ exponential tail of the correlator that is of interest.
For the mass calculation it is the small-$t$ short distance behavior
that is extracted via time-moments of $G(t)$, defined as:
\begin{equation}
G_{n,\text{latt}} = \sum_{t=0}^{T} (t / a)^{n} \, G(t)_{\text{latt}} \,.
\end{equation}

The time-moments $G_n$ have also been computed to N$^3$LO in 
perturbation theory~\cite{Chetyrkin:2006xg,Boughezal:2006px,Maier:2009fz}.
For $n \geq 4$,
\begin{equation} \label{G_n,pert}
G_{n,\text{pert}} = 
\frac{g_n(\a_{\MSbar}, \mu)}{(am_{h}(\mu))^{n-4}} \, .
\end{equation}
Here $m_h(\mu)$ is the $\MSbar$ quark mass at the scale $\mu$.
The basic strategy to extract the quark mass is
to compare $G_{n,\text{cont}}$, 
the continuum extrapolated $G_{n,\text{latt}}$ values,
with the perturbative expressions $G_{n,\text{pert}}$ in~\EQ{G_n,pert} 
(evaluated at a scale $\mu \sim m_h$),
and from these determine best-fit values for 
$\a_{\MSbar}(\mu)$ and $m_h(\mu)$.
For example, computing the continuum limit of $G_{4,\text{latt}}$
with physically tuned input charm masses $m_{0c}$,
one can obtain $\a_{\MSbar}(m_c)$, and then use this value in
$G_6$ to obtain $m_c(m_c)$.

The HPQCD collaboration carried out an analysis in~\cite{Chakraborty:2014aca} 
using reduced moments, $R_n$, 
which are simply related to the time-moments as
\begin{align}
R_4 &= G_4 / G_4^{(0)} \\
R_n &= \frac{1}{m_{0c}} (G_n / G_n^{(0)})^{1/(n-4)} \quad (n \geq 6) \,.
\label{R_n,latt}
\end{align}
where $G_n^{(0)}$ are the tree-level results for the moments.
Dividing by $G_n^{(0)}$ has the advantage of reducing lattice-spacing effects.
In continuum perturbation theory,
\begin{align}
R_4 &= r_{4}(\a_{\MSbar}, \mu) \label{R_4,pert}\\
R_n &= \frac{1}{m_c(\mu)} \, r_{n}(\a_{\MSbar}, \mu) \quad (n \geq 6) \, .
\label{R_n,pert}
\end{align}
Here $r_n$ are the perturbative expressions given by appropriate
powers of $g_n/g_n^{(0)}$, with $g_n^{(0)}$ the lowest order perturbative
result.
For a given $m_{0h}$ one computes the values of $R_n$
from \EQ{R_n,latt} and gets an estimate of $m_c(3m_h) = R_n/r_n(3m_h)$,
via \EQ{R_n,pert} (the scale $\mu$ was taken to be $3 m_h$).
In this way the scale dependence of $m_c^{\MSbar}(\mu)$ is determined.

\begin{figure}[htb]
\centering
\includegraphics[clip,height=3in]{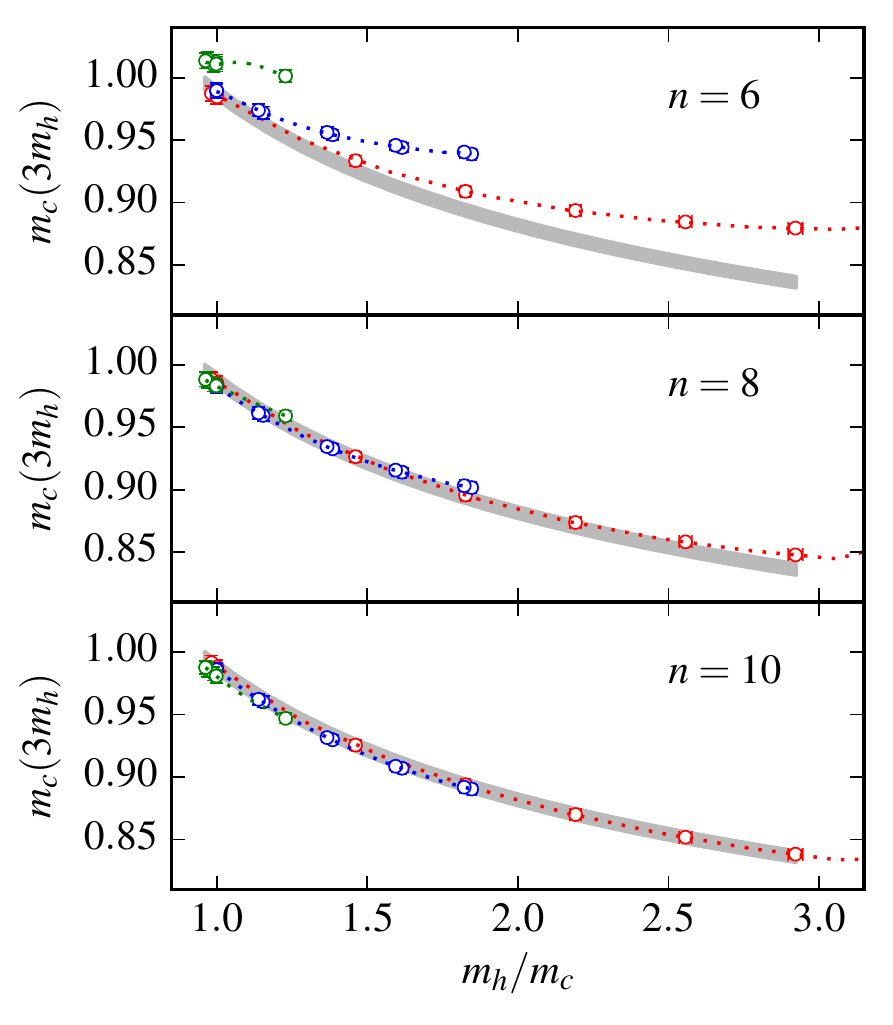}
\hskip 0.5cm
\includegraphics[clip,height=2.9in]{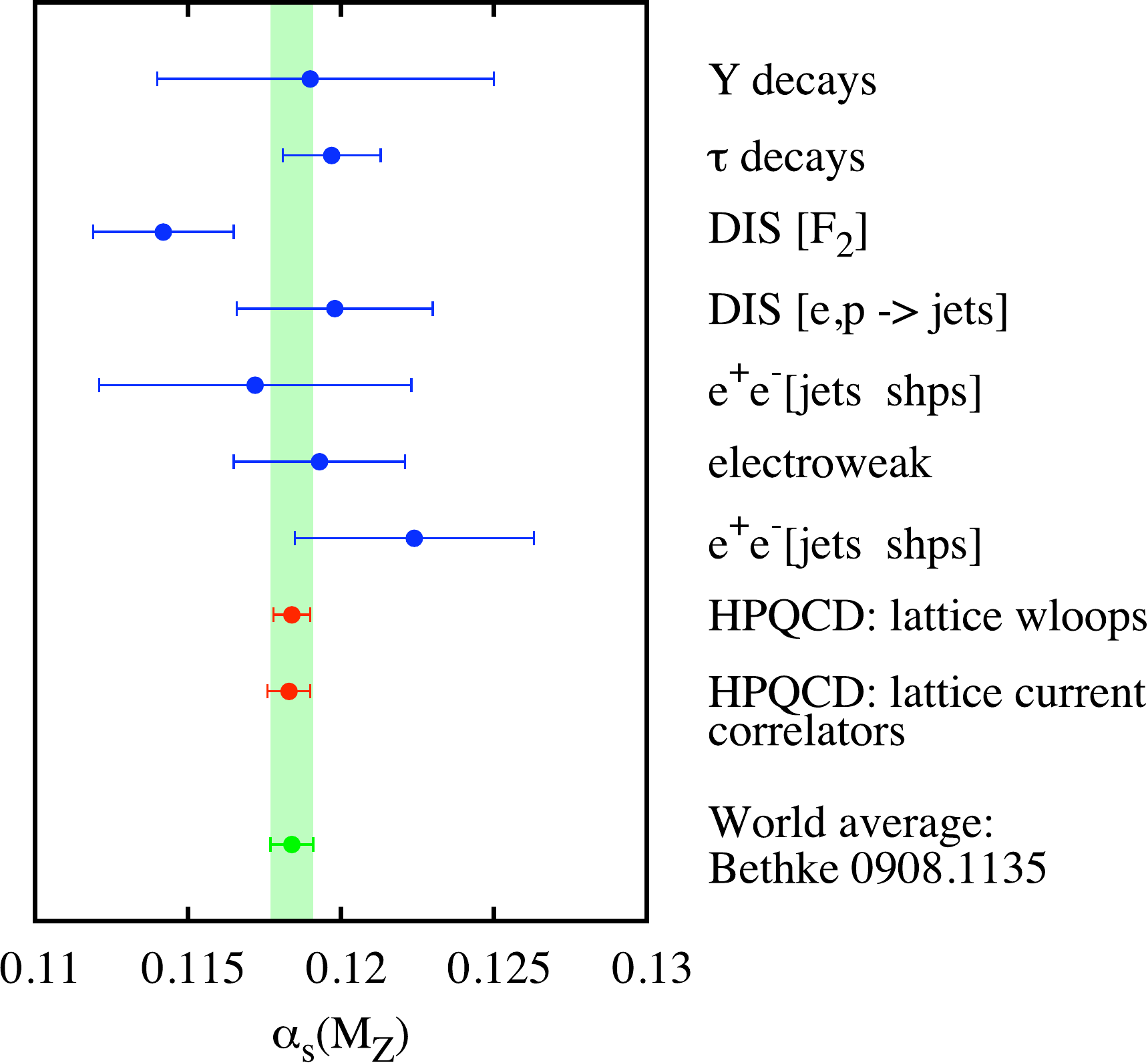}
\caption{
(Left) Data from~\cite{Chakraborty:2014aca} showing $m^{\MSbar}_c(\mu=3m_h)$
extracted from lattice data and perturbation theory for moments
$n=6,8,10$ using \EQ{R_n,pert}. The green/blue/red data points
correspond to lattice spacings of 0.12/0.09/0.06 fm.
The gray band shows the evolution of the best-fit value
for $m_c$ using perturbation theory.
(Right) Value of $\alpha_s^{\MSbar}(M_Z)$ 
from~\cite{Chakraborty:2014aca} compared with 
determinations based on various experimental inputs and a world average.
} 
\label{fig:mc-mh}
\end{figure}

The running of $m_c(\mu)$ was calculated this way 
in~\cite{Chakraborty:2014aca} using $n_f=2+1+1$ HISQ
ensembles.  The $n=4,6,8,10$ moments were computed using three
different lattice spacings $a \approx 0.12, 0.09, 0.06$ fm and for 
seven input masses from $m_h = m_c$ -- $0.7 m_b$.
The extractions of $m_c(3 m_h)$ from each of these
data points are shown in \FIG{mc-mh} (left),
along with the perturbative running.
\FIG{mc-mh} (right) shows the corresponding estimate of $\a_s$
extracted from this data, ran to $M_Z$ and compared with 
results based on other experimental inputs.

Estimates of $m_c(\mu)$ from time-moments
are subject to a number of
systematic uncertainties.  The truncation of perturbation theory
of course limits the precision. Fortunately the expansions of
$r_n = 1 + \sum_j \a^j r_{nj}$ are known for $j=1,2,3$ and $n \le 10$.
The lattice moments are sensitive to condensate terms 
not captured in the perturbative
expansions. These effects are suppressed like $(\Lambda_{\text{QCD}}/2m_h)^4$,
but they also grow with $n$.  The lattice data also has cut-off effects,
which grow like $\a_s \, (am_h)^2$ and decrease with increasing $n$; 
these trends are visible in \FIG{mc-mh}.

Fitting the moments data for $n=4,6,8,10$ to
Eqs.~\eqref{R_4,pert} and \eqref{R_n,pert}, 
HPQCD find
\begin{align}
m_{c}^{\MSbar}(3 \GeV, n_f=4) &= 0.9851(63) \GeV \\
\a_{s}^{\MSbar}(3 \GeV, n_f=4) &= 0.2545(37)  \, .
\end{align}

These are compatible with earlier 
$n_f=2+1$ results~\cite{McNeile:2010ji}.
The compatibility of $n+f=2+1$ and $n_f=2+1+1$ results suggests
that the effect of charm quarks in the sea can be treated perturbatively,
to this level of precision.

The JLQCD collaboration has recently utilized the current-current correlator
method with $n_f=2+1$ domain-wall fermions to determine
$m_c$ and $\a_s$~\cite{Nakayama:2015}.
Their calculation uses three lattices spacings 
$a \approx 0.08, 0.055, 0.044$ fm, and focuses on $R_6,R_8,$ and $R_{10}$,
from which they find
\begin{align}
m_{c}^{\MSbar}(3 \GeV, n_f=3) &= 0.9936(91) \GeV \\
\a_{s}^{\MSbar}(3 \GeV, n_f=3) &= 0.2526(92) \, .
\end{align}

\subsection{Comparison of results}
In~\cite{Carrasco:2014cwa} the ETMC collaboration 
use lattice RI/MOM techniques to determine
a mass renormalization factor $Z^{\text{RI}}_m(\mu, 1/a)$ connecting
the bare mass to the RI-scheme mass, 
$m_{c}^{\text{RI}}(\mu) = Z^{\text{RI}}_m(\mu, 1/a) \, m_{c0}$,  
which is converted
to the $\MSbar$ scheme using continuum perturbation theory.
Unlike the current-current correlator method, which uses a heavy input
mass to set the scale $\mu$, the RI/MOM calculation is extrapolated to
the chiral limit, and ETMC have generated mass degenerate $n_f=4$
ensembles for this purpose.  The $\chi$QCD collaboration have also
used RI/MOM methods for their $n_f = 2+1$ 
determination~\cite{Yang:2014sea,Liu:2013yxz}.

\begin{figure}
\centering
\hskip 2.5cm
\includegraphics[height=3in]{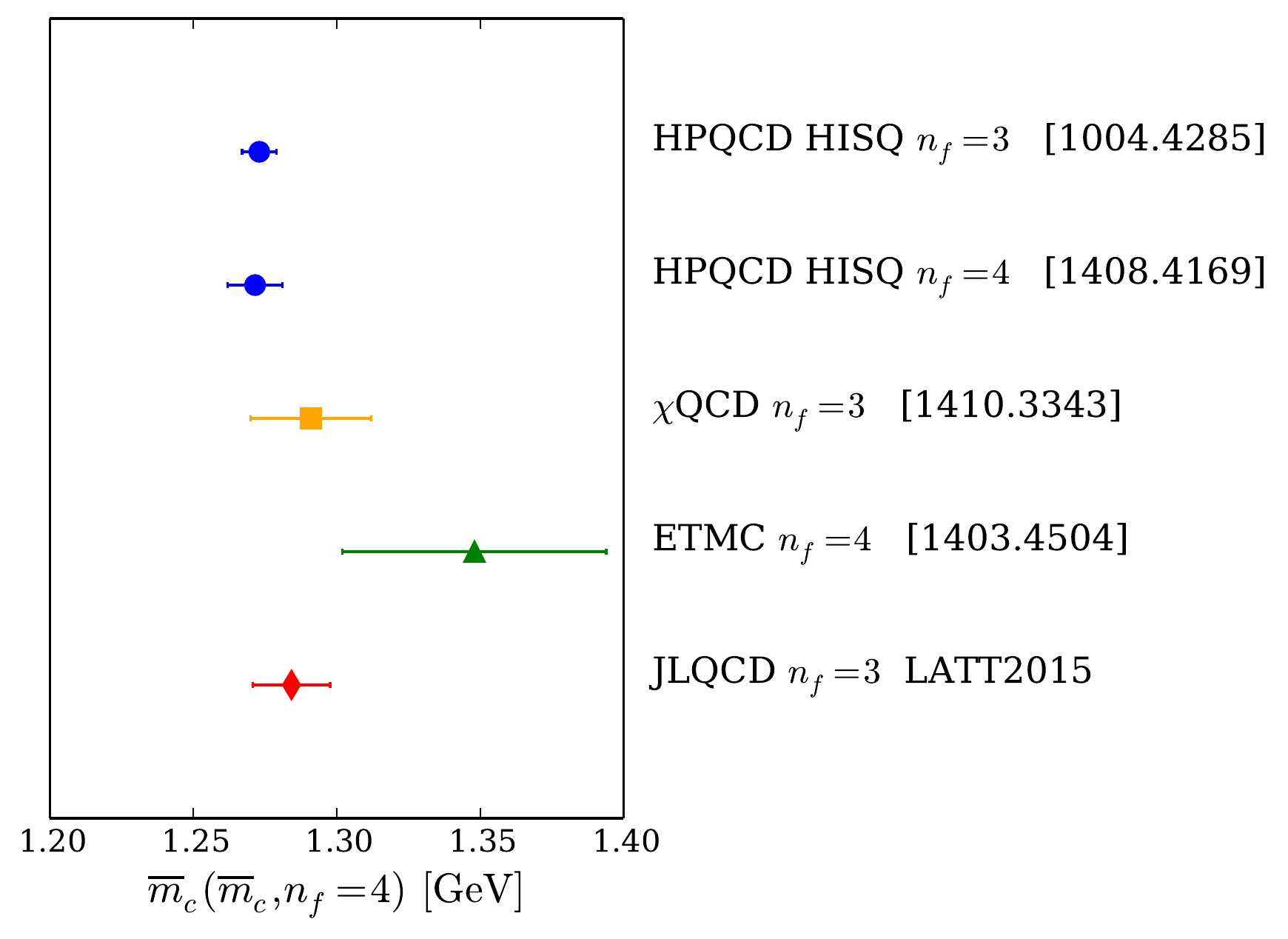}
\caption{
Comparison plot for determinations of $m_c^{\MSbar}(m_c^{\MSbar}, n_f=4)$,
computed from $n_f=2+1$ and $n_f=2+1+1$ simulations.
} \label{fig:mclatt}
\end{figure}

A comparison of recent lattice results for $m_{c}^{\MSbar}$
is shown in~\FIG{mclatt}.\\

\section{Bottom mass} \label{sec:bottom}
It is challenging to directly simulate the $b$ mass in relativistic
lattice simulations, since one would like $a m_{b0} \ll 1$ to keep
discretization effects under control.  Instead effective theories may
be employed such as non-relativistic QCD (NRQCD) or heavy-quark
effective theory (HQET).  It has recently become possible with 
improved relativistic actions to approach the $b$ mass,
making extrapolation methods viable.

\subsection{NRQCD approach}
The NRQCD Hamiltonian is written as an expansion in $v^2$, where $v$
is a typical velocity of a $b$ quark in the system of interest.
For example, $v^2 \sim 0.1$ in the $\Upsilon$ meson.
NRQCD calculations should be carried out with $a m_{b0} > 1$.
This has the advantage that the $b$ can be simulated using
relatively coarse lattices,
on the other hand it is less straightforward to extract continuum physics 
as compared to relativistic calculations.

The NRQCD current-correlator approach~\cite{Colquhoun:2014ica} 
is similar to the relativistic
approach described in \SEC{jj-charm}. One studies the time-moments of
Euclidean-time two-point correlators.  Unlike in the relativistic case,
here the currents need to be normalized,
\begin{equation}
J_{\mu}^{\text{NRQCD}} = Z_V J_{\mu}^{\text{cont}} \, .
\end{equation}
Then the time moments are related to continuum perturbation theory,
\begin{equation}
G^{\text{NRQCD}}_n = Z^{2}_{\text{V}} 
\frac{g_{n}(\a_{\MSbar},\mu)}{(am_b(\mu))^{n-2}} \, .
\end{equation}

\begin{figure}
\includegraphics[clip,width=0.49\textwidth]{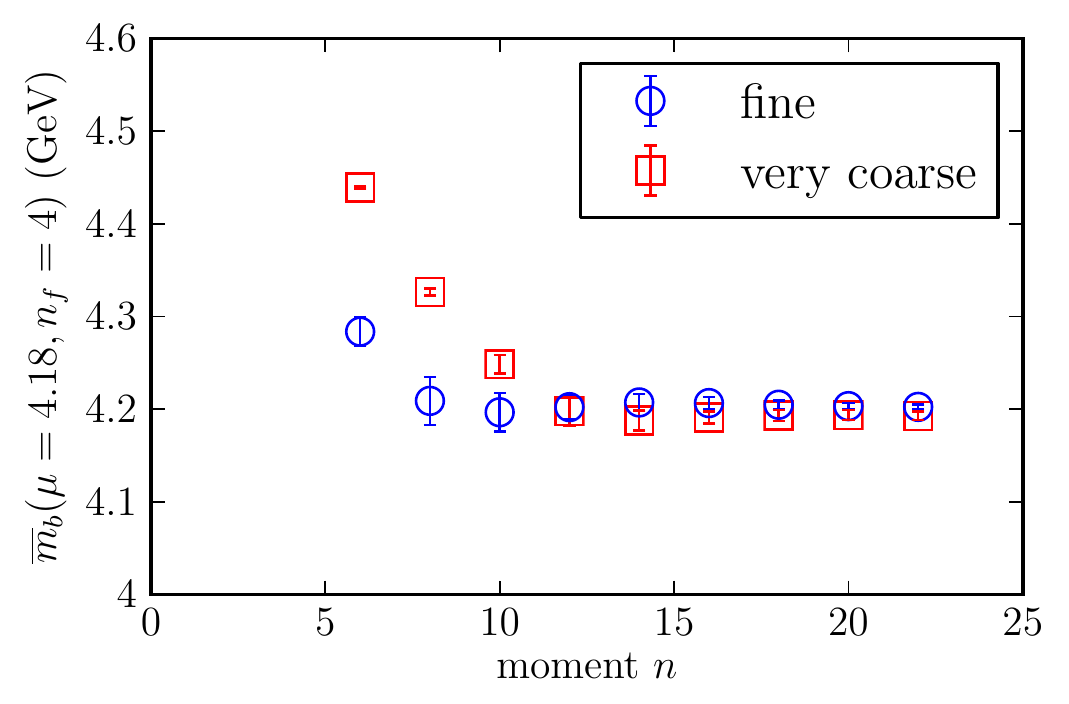}
\includegraphics[clip,width=0.49\textwidth]{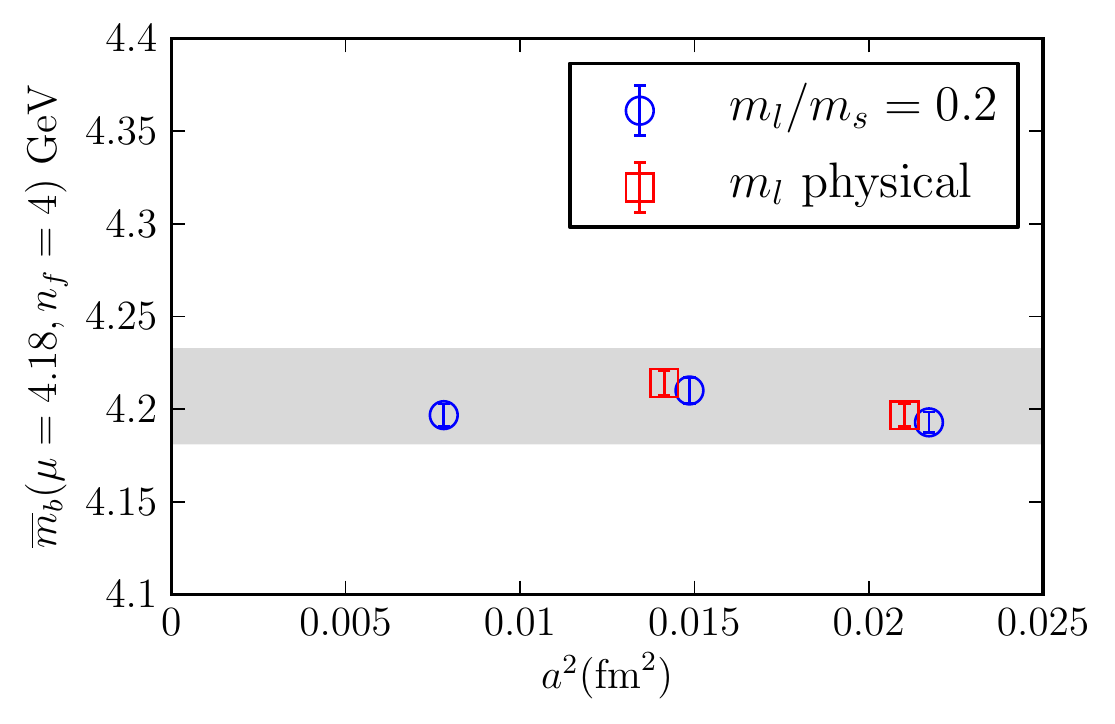}
\caption{
(Left) $m_{b}^{\MSbar}$ extracted from the moments of NRQCD
current-current correlators at two different lattice spacings
from~\cite{Colquhoun:2014ica}.
(Right) Results from the $n=18$ moment as a function of lattice
spacing and for two different light-quark masses.  The gray
band gives the continuum determination with the total error budget.
} \label{fig:moments-mb}
\end{figure}

Constructing ratios of successive moments, the factors of $Z_V$ 
can be canceled.  Because the continuum limit cannot be approached
directly one instead studies  
$m_{b}$ as a function of the moment number.
Compared to the charm case, condensate contributions which 
grow with moment number are more suppressed at
the heavier quark mass.
A ``plateau'' in $m_b$ as a function of moment number implies that $n$
is sufficiently large for discretization effects to be small.
Such a plateau from~\cite{Colquhoun:2014ica}
is shown in~\FIG{moments-mb} (left).

Results at three lattice spacings and with two different light-quark
masses for $n=18$ are shown in~\FIG{moments-mb}~(right).
A fit to this data, including systematic errors, and perturbatively evolved
to $m_b$ gives
\begin{equation}
m_b^{\MSbar}(m_b^{\MSbar}, n_f=5)  = 4.196(23) \text{ GeV} \,.
\end{equation}

\begin{figure}
\centering
\hskip 2.5cm
\includegraphics[height=3in]{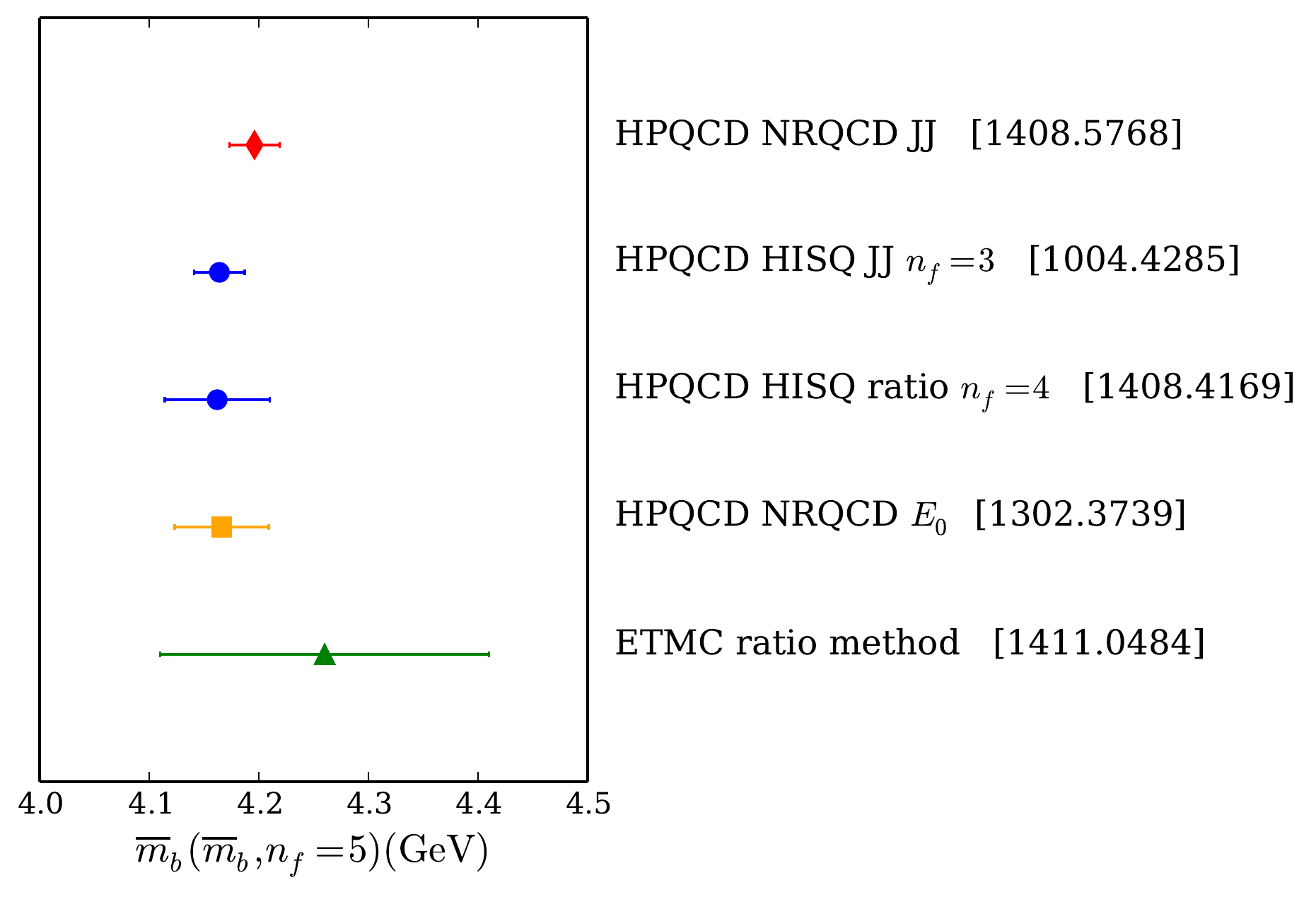}
\caption{
Comparison plot for determinations of $m_b^{\MSbar}(m_b, n_f=5)$,
computed from $n_f=2+1$ and $n_f=2+1+1$ simulations.
} \label{fig:mblatt_update}
\end{figure}

This result is compared with others in~\FIG{mblatt_update}.
It is significant that the values in the figure are calculated
using a range of techniques. In~\cite{McNeile:2010ji} results are extrapolated
to $m_b$ from below, using a relativistic action as described in
\SEC{jj-charm}.  This calculation is based on a different range of moment
numbers, and uses a different action than~\cite{Colquhoun:2014ica}.
The work of~\cite{Lee:2013mla} uses the binding energy of $\Upsilon$
and $B_s$ mesons, computed using NRQCD and lattice perturbation
theory, to determine the heavy quark pole mass, which is then
converted to the $\MSbar$ mass with continuum perturbation theory.

\subsection{Ratio method}
The ETMC collaboration have  
used the 
\emph{ratio method}~\cite{Blossier:2009hg} to 
extrapolate relativistic $n_f = 2+1+1$
simulation results around the charm mass 
to the bottom mass~\cite{Bussone:2014cha}.
The method is based on the expectation from HQET that
\begin{equation} \label{static-limit}
\lim_{m_h^{\text{pole}}\rightarrow \infty} 
\frac{M_{hl}}{m_h^{\text{pole}}} = \text{constant} \,,
\end{equation}
where $M_{hl}$ is the mass of a heavy-light meson and
$m_{h}^{\text{pole}}$ is the heavy quark pole mass.

They use simulation data consisting of ratios of meson masses,
$M_{hl}(m_h)/M_{hl}(m_h/\lambda)$,
computed for a
series of masses $m_h$ around the charm mass, e.g.:
$m_{h}^{(0)} = m_c$, $m_{h}^{(1)} =\lambda m_c$, ..., 
$m_{h}^{(n)} =\lambda^n m_c$.
These ratios have the advantage that discretization effects proportional
to $(am_h)^2$ are largely canceled.  
From this data they construct the function
\begin{equation} \label{y(m_h)}
y(m_h, \lambda) = \lambda^{-1} \frac{M_{hl}(m_h)}{M_{hl}(m_h/\lambda)}
\frac{\rho(m_h/\lambda)}{\rho(m_h)} \, .
\end{equation}
The functions $\rho(m_h)$ on the r.h.s.\ of~\EQ{y(m_h)} relate 
the pole mass to the $\MSbar$ mass and are known
to $\text{N}^3$LO in perturbation theory.
$y(m_h, \lambda)$ satisfies 
$\lim_{m_h \rightarrow \infty} y(m_h,\lambda) = 1$
on account of~\EQ{static-limit}, and so its value can interpolated 
between the charm region and the static limit using a motivated fit ansatz.
Rewriting~\EQ{y(m_h)}, the combination
$\lambda \, y(m_h, \lambda) \frac{\rho(m_h)}{\rho(m_h/\lambda)}$
is then a known function that evolves $M_{hl}(m_h/\lambda)$ to $M_{hl}(m_h)$.
Choosing $\lambda$ such that $M_{hl}(m_{h}^{(N)}) = M^{\text{phys}}_{bl}$
for some $N$, they determine the $b$ mass from $m_b = \lambda^N m_c$.

\section{Mass ratios} \label{sec:ratios}
Mass parameters are inputs to lattice QCD simulations,
these are pure numbers $(a m_0)$ corresponding to masses expressed
in units of the lattice spacing $a$.
There is one bare mass parameter for each quark in the simulation,
and these
must be tuned to reproduce the physics of QCD.  
The bare mass parameters are tuned by measuring low-energy observables
such as meson masses, and requiring that these are be equal to their physical
values.
After this set of observables has been used to tune the simulation parameters,
one has a set of numbers $\{(am_{ud0}), (am_{s0}), (am_{c0}) \}$.
The bare lattice inputs are defined at the cutoff scale and depend on
the details of the discretization.
However, ratios of input masses are equal to the ratios of $\MSbar$ masses,
up to discretization effects that vanish in the continuum,
\begin{equation}
\frac{a m1_0}{a m2_0} = \frac{m1^{\MSbar}(\mu)}{m2^{\MSbar}(\mu)} + \O(a^2) \, .
\end{equation}
Thus once the $\MSbar$ mass is known for one quark in the theory,
this can be converted to the $\MSbar$ masses for the others using
the input mass parameters.

\begin{figure}[htb]
\includegraphics[clip,height=2in]{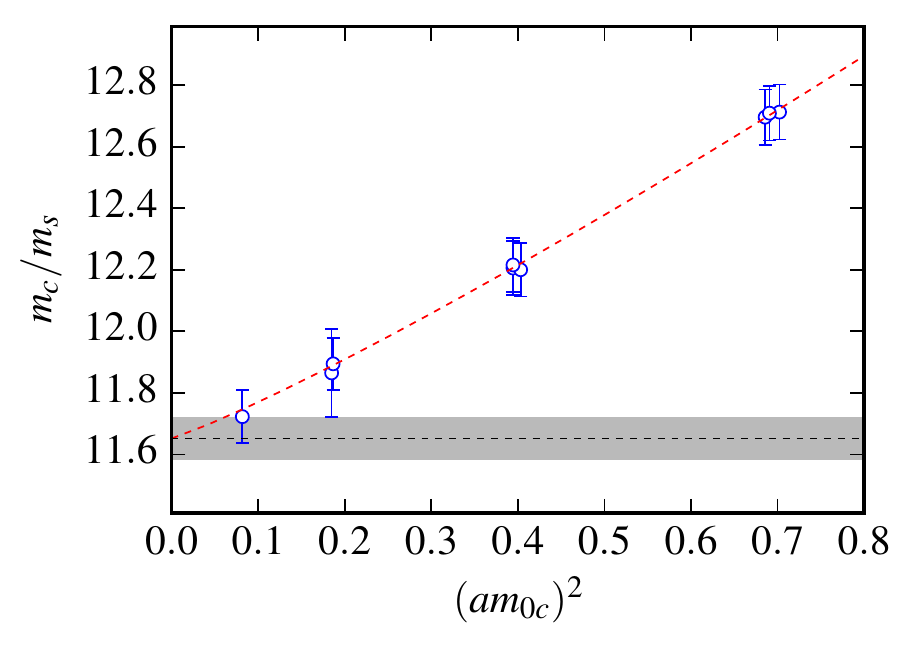}
\includegraphics[clip,height=2in]{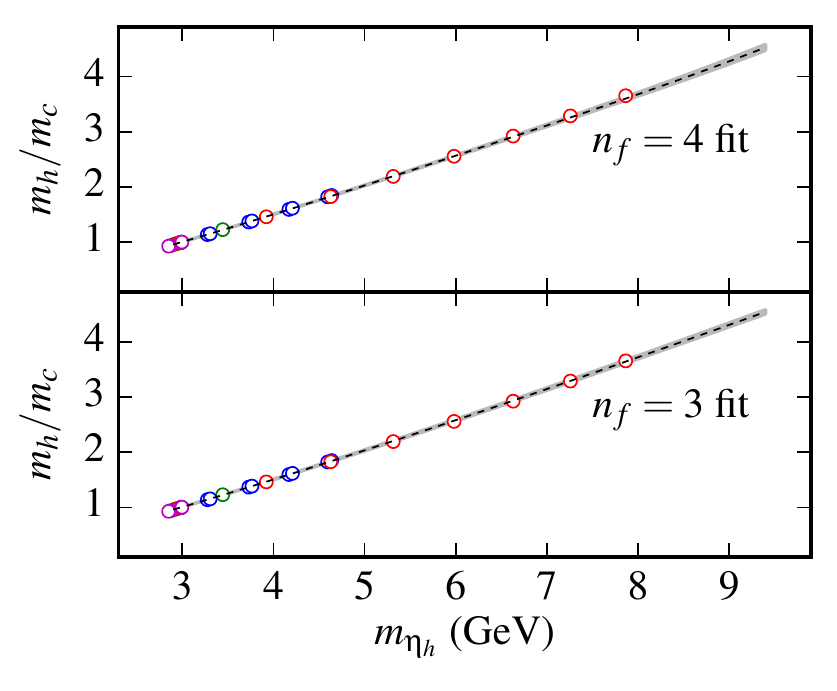}
\caption{
(Left) Continuum extrapolation of the bare quark mass ratio
$\frac{m_{0c}}{m_{0s}}$ from~\cite{Chakraborty:2014aca}.
(Right) Extrapolation of $\frac{m_{0h}}{m_{0c}}$ in $m_{\eta_h}$ to $m_{\eta_b}$ using
simulation data from~\cite{Chakraborty:2014aca}.
Magenta/green/blue/red points are from lattice spacings of
0.15/0.12/0.09/0.06 fm.
} \label{fig:mc-ms}
\end{figure}

An example of this is shown in~\FIG{mc-ms} (left), for the input ratio
$m_{0c}/m_{0s}$ from~\cite{Chakraborty:2014aca}. 
In the continuum HPQCD find that
\begin{equation}
\frac{m_c(\mu, n_f)}{m_s(\mu, n_f)} = 11.652(65)
\end{equation}
Using their result for $m_{c}^{\MSbar}(\mu)$ from the current-current
correlator method discussed in~\SEC{jj-charm},
they obtain
\begin{equation}
m_{s}^{\MSbar}(3 \GeV, n_f=3) = 84.7(7) \MeV \, ,
\end{equation}
which is the most precise estimate to date.
\FIG{mc-ms} (right) shows a result from~\cite{Chakraborty:2014aca}
using input mass ratios to obtain $m_b^{\MSbar}$.  Here the
input mass is increased from $m_{c0}$ towards $m_{b0}$,
and finally an extrapolation performed to obtain
\begin{equation}
\frac{m_b(\mu, n_f)}{m_c(\mu, n_f)} = 4.528(54)
\end{equation}
\begin{equation}
m_{b}^{\MSbar}(m_b, n_f=5) = 4.162(48) \GeV
\end{equation}

\section{Conclusion} \label{sec:conclusion}
Recent progress in lattice determinations of charm and bottom
quark mass was reviewed.  
In order to achieve (sub-)percent level uncertainties
for these quantities, it is important that determinations come both
from a variety of calculational strategies, and via independent
measurements from different groups.

The most precise quoted values for $c$ mass
presently come from calculations of current-current correlators,
comparing these to perturbation theory,
where a heavy ($\sim m_c$) input mass sets the scale $\mu$.
The precision in the value of the charm mass
can be cascaded to the other masses using bare quark-mass ratios, which
are determined in the tuning of simulation parameters to their physical values.

Calculations of $b$ mass are done either using an effective-theory framework
for the $b$ quark or extrapolating relativistic simulations from lower-mass region
where discretization effects are under control. Extrapolation methods
will continue to improve as ensembles with smaller lattice spacings
become available.
First steps have been taken towards a fully relativistic treatment of
the $b$ quark~\cite{Chakraborty:2014aca,McNeile:2010ji}.  
This will lead not only to more precise
values for the $b$ itself, but through the use of mass ratios
should improve determinations of the other quark masses as well.

\Acknowledgements
I would like to thank the organizers of Charm 2015 for a 
very enjoyable conference,
and the participants for many illuminating discussions,
in particular M.~Padmanath, Sasa Prelovsek, and Vicent Mateu.
I would like to thank Christine Davies, Yi-Bo Yang, Petros Dimopoulos, 
and Katsumasa Nakayama for providing material for this review,
and Christine Davies for providing feedback on the manuscript.


\end{document}